\documentclass[acmlarge,screen]{acmart}

\AtBeginDocument{%
  }

\setcopyright{acmcopyright}
\copyrightyear{2023}
\acmYear{2023}
\acmDOI{XXXXXXX.XXXXXXX}

\acmJournal{JACM}
\acmVolume{37}
\acmNumber{4}
\acmArticle{111}
\acmMonth{8}

\usepackage{natbib}
\usepackage{xcolor}
\usepackage{amsmath}
\usepackage{enumitem}
\usepackage{wrapfig}
\usepackage{xltabular}
\usepackage[utf8]{inputenc}
\usepackage{tikz}

\usepackage{array}
\usepackage{makecell}
\usepackage{multirow}
\usepackage[acronym]{glossaries}
\usepackage{color,soul}
\soulregister\acrshort7
\soulregister\acrfull7

\newtheorem{hyp}{Hypothesis}

\begin{document}

\title[Proc. ACM Interact. Mob. Wearable Ubiquitous Technol.]{Ongoing Tracking of Engagement in Motor Learning}


\author{Segev Shlomov}
\authornote{All authors contributed equally to this research.}
\email{segev.shlomov1@ibm.com}
\author{Nitzan Guetta}
\authornotemark[2]
\email{nitzangu@post.bgu.ac.il}
\author{Jonathan Muehlstein}
\authornotemark[2]
\email{jon.muehlst@gmail.com}
\author{Lior Limonad}
\authornotemark[2]
\email{liorli@ibm.com}
\affiliation{%
  \institution{IBM Research - Haifa}
  %\streetaddress{P.O. Box 1212}
  \country{Israel}
}


\begin{abstract}
Teaching motor skills such as playing music, handwriting, and driving, can greatly benefit from recently developed technologies such as wearable gloves for haptic feedback or robotic sensorimotor exoskeletons for the mediation of effective human-human and robot-human physical interactions. At the heart of such teacher-learner interactions still stands the critical role of the ongoing feedback a teacher can get about the student’s engagement state during the learning and practice sessions. 
Particularly for motor learning, such feedback is an essential functionality in a system that is developed to guide a teacher on how to control the intensity of the physical interaction, and to best adapt it to the gradually evolving performance of the learner. In this paper, our focus is on the development of a near real-time machine-learning model that can acquire its input from a set of readily available, noninvasive, privacy-preserving, body-worn sensors, for the benefit of tracking the engagement of the learner in the motor task. We used the specific case of violin playing as a target domain in which data were empirically acquired, the latent construct of engagement in motor learning was carefully developed for data labeling, and a machine-learning model was rigorously trained and validated.
\end{abstract}




\received{15 August 2023}
\received[revised]{15 August 2023}
\received[accepted]{15 August 2023}

\newacronym{imu}{IMU}{Inertial Motion Unit}
\newacronym{gsr}{GSR}{Galvanic Skin Response}
\newacronym{emg}{EMG}{Electromyography}
\newacronym{eeg}{EEG}{Electroencephalogram}
\newacronym{ecg}{ECG}{Electrocardiography}
\newacronym{gyr}{GYR}{Gyroscope}
\newacronym{hrv}{HRV}{Heart-rate Variability}
\newacronym{ans}{ANS}{Autonomous Nervous System}
\newacronym{cns}{CNS}{Central Nervous System}
\newacronym{ml}{ML}{Machine Learning}
\newacronym{gmsi}{Gold-MSI}{Goldsmiths Musical Sophistication Index}
\newacronym{stai}{STAI}{State-Trait Anxiety Inventory}
\newacronym{crs}{CRS}{Comfort Rating Scale}
\newacronym{eng}{EML}{Engagement in Motor-learning}
\newacronym{ucbm}{UCBM}{Università Campus Bio-Medico di Roma}
\maketitle

\section{Introduction}
Our focus in this work is part of an effort to develop innovative instrumentation, which combines hardware and software, with the purpose of physically coupling humans to enhance complex motor skill learning, such as handwriting and music learning.
The work is conducted under a project that develops an exoskeleton-based human-human and robot-human interactions for motor learning. Learning to play the violin, as one specific example of a motor skill, is one of the most complex musical instruments to learn~\cite{nicola-2021-violin}. Such learning relies on physical interactions in which the learner continuously employs sensorimotor haptics to learn from others~\cite{Basdogan2000,Reed2008,ganesh2014two}. The effectiveness of the learner is driven by a variety of factors. Particularly,% in this work
 we attend to the desire of the teacher to be able to adapt their teaching pace and level of difficulty, while continuously ensuring that the learner is in the `right' mental state, an aspect that significantly influences the learning process and the eventual successful completion of a task~\cite{bower1992,martocchio1994effects,warr1995trainee}. This can be achieved by utilizing wearable sensors as input data to identify users' mental and emotional status and change the engagement platform's behavior accordingly. Recent work on human activity and mental state recognition (e.g., fatigue, stress) has most of its focus on leveraging body-worn measurements and extracted features such as gait and breathing patterns using \acrfullpl{imu}, \acrfull{gyr}, \acrfull{hrv}, and \acrfull{gsr}~\cite{shaffer2017overview}. Based on such measures, machine-learning models provide insights regarding the physical, mental, and emotional states of the student. Specifically, in this work, our goal is a rigorous methodological development of a classification model for near real-time determination of learners' engagement in the performance of a motor task, namely \acrfull{eng}. We also demonstrate how observational sensor data can be integrated with adequate labeling which is essential for the employment of supervised machine learning. % via careful development of a behavioral model that is tailored for the capturing of \acrshort{eng} as its latent construct utilizing self reported scales from violin players. 
Such labeling was achieved via a behavioral model that was rigorously developed for capturing \acrshort{eng} as its latent construct utilizing self-reported scales from violin players.
The developed model has its theoretic foundations rooted in the Flow theory~\cite{csikszentmihalyi1990flow}. Subsequently, data that were gathered with the behavioral model were further interpolated in the context of each individual excerpt to increase its density and enable the use of supervised \acrshort{ml}. This interpolation also promotes the responsiveness of the developed \acrshort{ml} model, where we also ensured to not synthetically over-inflate the accuracy of the developed model. Finally, we also examined the performance of the model with respect to subsets of sensor input. 

Ultimately, the resulting model developed here will be embedded for ongoing feedback to teachers, and for the control of a modular robotic platform that integrates wearable technology as a feedback mechanism. This integration aims to facilitate more effective human-human and human-robot interaction, driving motor learning activities such as playing music or handwriting. Overall, the result of this work enables the monitoring of the student's engagement state as a second modality that complements the conventional output of the activity, whether it is vocal as in the case of music playing, or graphical as in the case of handwriting.

\section{Related Literature}
\label{sec:background}
Recent advances with body-worn sensors have the potential to capture data that may cater to the monitoring of a wide variety of physiological, mental, and emotional conditions. Contemporary machine learning techniques are also well-developed to capture complex, multi-tiered correspondences between such sensor data and mental conditions. Thus, it is essential for the training of such models to combine both, sensor data relevant for the classification task input, and also labeling of the data with the target states to be predicted. It is key to have such data with both ingredients put in sync in our training set for the construction of an accurate classifier.%Lior: may drop last sensence

There is a vast literature on the \emph{engagement} concept captured via single or multi-sensor instrumentation, with subclassifications split between cognitive and emotional. The concept itself is somewhat vaguely defined, in many cases due to being task-oriented in nature. Prior work features different target activities such as in-class learning~\cite{3550335, 3550328, 7033174}, learning with VR~\cite{DUBOVI2022104495}, internet web browsing~\cite{bulger-2008}, maker learning~\cite{Lee2019AWA}, general task engagement~\cite{5349483}, and even military operational duties~\cite{Berka-2007}. A recent exhaustive survey~\cite{Shan-2021} summarizes a plethora of methods for cognitive engagement measurements, including self-report scales, observations, interviews, teacher ratings, experience sampling, eye-tracking, physiological sensors, trace analysis, and content analysis. Specifically, with respect to sensing devices employed for its measurement, \acrshort{gsr} and \acrfull{emg} are determined to be most common for emotional engagement, while \acrfull{eeg} is commonly used for cognitive engagement. \acrshort{eeg} is a good solution for highly reliable ongoing measurements, as demonstrated in~\cite{Berka-2007}, but it is very tedious to employ. Alternative efforts include less invasive means such as facial expression, eye tracking, \acrshort{gsr}, and body posture captured with cameras (e.g.,~\cite{DUBOVI2022104495,7033174}), sometimes attempted as a surrogate for \acrshort{imu} motion sensing~\cite{3596261}.

Different measurements are relevant for the capturing of \acrshort{eng} as a specific type of cognitive and emotional manifestation, based on different physiological parameters associated with the \acrfull{cns}, and with changes of the \acrfull{ans}. Such activity can be analyzed by monitoring electrodermal, respiratory, cardiac, and inertial activities. %Electrodermal activity is a sensitive peripheral index of the sympathetic division of the ANS and can be appreciated by measuring changes in skin conductance. Gloves integrating textile electrodes for 
Galvanic Skin Response (\acrshort{gsr}) analysis has been developed to discriminate with good accuracy different levels of arousal~\cite{lanata2012eda}. %Recently, researchers have started investigating the 
Heart Rate Variability (\acrshort{hrv}) in both time and frequency domains can provide a noninvasive assessment of \acrshort{ans} function~\cite{Acharya2006heart}. The analysis of \acrshort{hrv} nonlinear measures has been found to be important for emotional arousal and valence recognition from \acrshort{ans} signals~\cite{valenza2014,katsis2011}. Similarly, cessation of respiration tracking (e.g., using \acrshortpl{imu}) can be associated with stress~\cite{gorman2004,wu2012}.

% Our work had to adhere to specific requirements: we needed to accommodate for the unique activity of \emph{motor learning} as the target activity. 
In our instrumental design, we had to meet specific requirements and accommodate the unique activity of \emph{motor learning} as our target. % to be engaged with. 
Hence, we had to rule out frequent interruptions during the main activity. %We also had to ensure that we employ readily available, privacy-preserving sensors, conforming to our intentions to embed them within a broader interaction training platform to be deployed in schools or industrial training facilities, where privacy is a strict concern. 
We also had to ensure that we used readily available, privacy-preserving sensors. This conformed to our intentions to embed them within a broader interaction training platform, destined for deployment in schools or industrial training facilities where privacy is a paramount concern. Our sensor selection also aimed to have a minimal footprint, aiming to minimize discomfort during task performances such as violin playing.

\begin{figure}\centering
\includegraphics[width=0.6\linewidth]{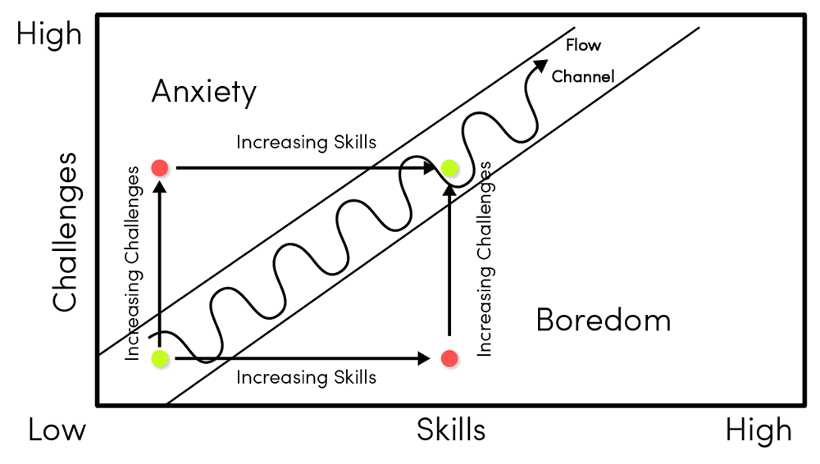}
\caption{ The Flow theory: `flow' is attained when the level of skill matches the perceived complexity of the challenge.}
\label{fig:flow-theory}
\end{figure}

%For the latter element of capturing the state of interest to be predicted, 
Our aim %was 
to measure the condition of being engaged in the performance of a motor task was concretely instantiated by the activity of playing the violin during practice sessions in this work. For this, we adopt from the theory of \emph{optimal experience} or `Flow' by Csikszentmihalyi~\cite{csikszentmihalyi1990flow}, which conceptualizes flow as a state of concentration that amounts to absolute absorption in an activity. It refers to ``the state in which people are so involved in an activity that nothing else seems to matter.'' Henceforth, matching our concept of \acrshort{eng}. Specific to the domain of music, the theory also attributes music itself as auditory information that helps ``organize the mind...listening to music wards off boredom and anxiety, and when seriously attended to, it can induce flow experiences...even greater rewards are open to those who learn to make music...'' Aside from aiding with the conceptualization of the notion itself, the same theory also guided us in how to design an experimental setup that actively induces the conditions for \acrshort{eng}. 
% This stems from its model of Challenges $\&$ Skills~\cite{csikszentmihalyi1988optimal} according to which the conditions for optimal experience are attained when balancing between the \textit{challenge perceived} in a given situation and \textit{the skills} a person brings to it, as illustrated in Figure~\ref{fig:flow-theory}. 
This is based on its model of `Challenges-$\&$-Skills'~\cite{csikszentmihalyi1988optimal}. According to this model, conditions for optimal experience are achieved when there's a balance between the challenge perceived in a given situation and the skills a person possesses, as illustrated in Figure~\ref{fig:flow-theory}.
We also needed to attend to the dynamic nature of \acrshort{eng} as ``to remain in flow, one must increase the complexity of the activity by developing new skills and taking on new challenges.''~\cite{csikszentmihalyi1988optimal} Hence, conditions to induce \acrshort{eng} are different between people having evolving levels of expertise, via learning and practicing, which requires corresponding change %and manipulation 
in the difficulty level of the task at hand. Only when in a proper balance, a person is freed from interrupting questions such as `am I doing well?'~\cite{csikszentmihalyi2000beyond}.

\section{Method}
\label{sec:method}

\begin{figure}[ht]
\centering
\includegraphics[width=0.95\linewidth]{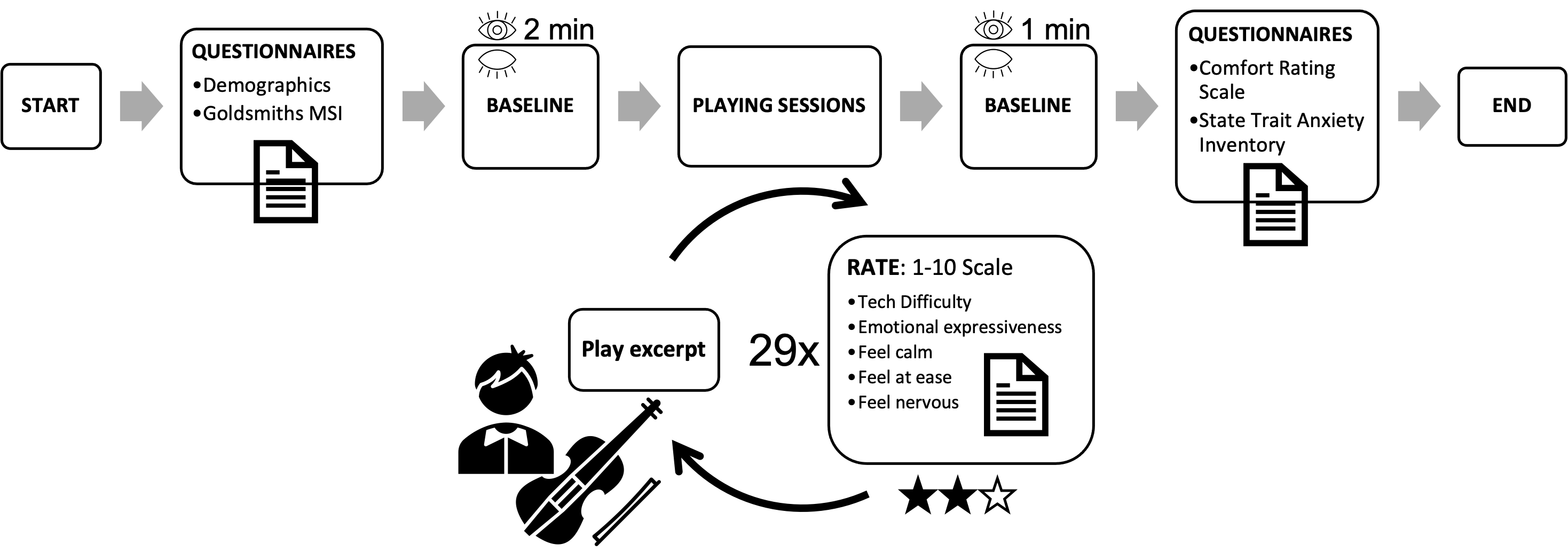}
\caption{Experimental protocol}
\label{experiment-protocol}
\end{figure}

We pursued a two-staged \acrfull{ml} model development approach. Our first stage comprises a gradual empirical development of an underlying behavioral model that serves as a basis for the capturing of the main target mental state of \acrshort{eng} as a latent construct. 
% Guided by the Challenges-$\&$-Skills view in the theory of flow, our experimental protocol was configured %to trigger a designated manipulation that is intended 
% to alter the conditions that are needed for the emergence of \acrshort{eng}.
Guided by the Challenges-$\&$-Skills view in the theory of flow, our experimental protocol was designed to modify the conditions necessary for the emergence of \acrshort{eng}.
As a concrete operationalization, 
we employed the task of violin playing with skilled musicians during practice sessions while manipulating the level of difficulty of the tasks. The behavioral model was rigorously built via inspection of the inter-construct relationships in the model and statistical assessment of their significance.
% Self-reported scales by the musicians along the experimental tasks were used as data input, complemented by synchronized sensor recordings. The results from the first stage provided us with a dataset that integrated time series sensor data with validated labeling of \acrshort{eng}.
Self-reported scales by the musicians during the experimental tasks were utilized as data input, complemented by synchronized sensor recordings. The results from the initial stage provided us with a dataset that incorporated time series sensor data along with validated labeling of \acrshort{eng}.

% As a second stage, we delved into the analysis of the dataset captured and its enrichment with corresponding features, for the elicitation of a near-real-time \acrshort{ml} algorithm that can leverage only the sensory input as a means to determine the target condition of \acrshort{eng}. 
In the second stage, we delved into the analysis of the captured dataset and enriched it with corresponding features. This was done for the elicitation of a near-real-time \acrshort{ml} algorithm that relies solely on sensory input to determine the target condition of \acrshort{eng}.
This means using the raw labeled dataset, enriching it with corresponding features, and developing a machine-learning 
model to classify the perceived level of \acrshort{eng} via sensor data as a sufficiently accurate surrogate for the paper-and-pencil scales that were employed in the first stage.
% This %second stage also 
% was followed by careful assessment of a handful of specific \acrshort{ml} models, and selection of the one that presented a good combination of accuracy and responsiveness using conventional metrics. For pragmatic purposes, we also complemented this latter effort with sensor sensitivity analysis, to determine the importance of individual sensor types in the predictions. 
This was followed by a meticulous evaluation of a few specific \acrshort{ml} models, and the selection of the one that exhibited a favorable balance between accuracy and responsiveness, as assessed using conventional metrics. To address practical considerations, we additionally augmented this endeavor with sensor sensitivity analysis to ascertain the significance of individual sensor types in the predictive outcomes.

\section{Stage I: The Experiment}\label{sec:experiment}

The overall experimental process that was pursued is illustrated in Figure~\ref{experiment-protocol}. Nine adult musicians (3 Males; 6 Females), with a mean age of 24 $\pm$ 2 years, and holding a conservatory degree in Violin, were randomly recruited and volunteered to participate in our study.

Ethical board approval was granted,
% by \acrshort{ucbm} 
and informed consent was obtained from all participants. As a first step, each participating violinist in the experiment was requested to complete a form containing general demographic information (e.g., gender, age, nationality). The questionnaire also included a section with the previously validated \acrfull{gmsi} scale~\cite{mullensiefen2014measuring}. This scale consists of 39 questions designed to assess professional musical expertise across various dimensions of musicality. % such as: musical training, musical engagement in general, perceptual abilities, and emotions. 
Particularly, the musician's musical profile according to their \acrshort{gmsi} facet scores (average and percentile) catered: Active Engagement (47.14, 69), Perceptual Abilities (51.86, 55), Musical Training (41.14, 89), Emotions (33.29, 38), Singing Abilities (34.86, 61), and General Sophistication (96.86, 74). In comparison to the population norm, our subjects exhibited an average musical training level that ranked in the $89th$ percentile.

\begin{figure}
\centering
\includegraphics[width=0.7\linewidth]{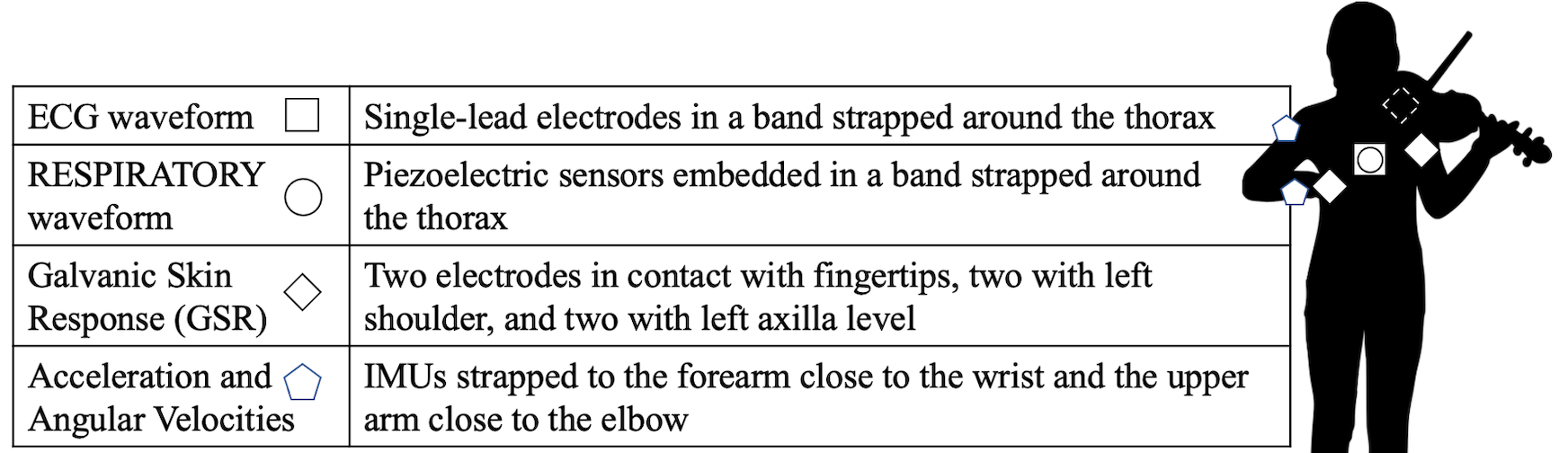}
\caption{Wearable sensors}
\label{wearable-sensors}
\end{figure}

After filling out the questionnaires, the participants were equipped with a set of body-worn sensory devices, as illustrated in Figure~\ref{wearable-sensors}. This included measurements of \acrfull{ecg}, respiration rate, skin conductance (\acrshort{gsr}), and motion tracking (\acrshort{imu}) of the wrist and the upper arm. The \acrshort{gsr} Shimmer sensor was positioned on the index and middle finger of the bow-holding hand. Subsequently, a baseline measurement was taken in a relatively stationary standing position, capturing two minutes with eyes shut and two minutes with eyes open. These recordings were conducted to facilitate data normalization during the later analysis phase.

% \begin{figure}
% \centering
% \includegraphics[width=0.5\textwidth]{figures/sensors-ibm-cut.png}
% \caption{Wearable sensors}
% \label{wearable-sensors}
% \end{figure}

Following the baseline step, the main part of the experimental protocol was initiated. This consisted of repetitive playing sessions during which each violinist was instructed to play a series of 29 musical excerpts in a randomly assigned order. These excerpts constitute the total set manifested by a combination of excerpt \emph{category} and \emph{tempo} \cite{fernandez2016influence,liu2018effects} as further elaborated in Section~\ref{sec: behave-model-dev}.

% After playing each individual excerpt, a short questionnaire was administered in which the violinists were asked to rate their perceptions of the technical difficulty, the emotional expressiveness, how much they felt being calm, at ease, nervous, and the level of discomfort they have felt as a result of wearing the sensors while playing the violin. 
After playing each individual excerpt, a questionnaire was administered to the violinists, asking them to rate their perceptions of the technical difficulty, emotional expressiveness, their level of calmness, ease, nervousness, and discomfort \cite{knight2002comfort} resulting from wearing the sensors while playing the violin.
All scales have been the basis for the behavioral model development in Section~\ref{sec: behave-model-dev}. The former two items have been part of the constructs that capture the intended manipulation of challenge complexity, and the latter were adopted from the %full 
scales %at the end of the experiment.
that were administered at the end of the experiment.

Concluding the experimental protocol, each violinist repeated two 1-minute lasting baselines (with eyes open and eyes closed). Finally, once the sensor devices were removed, the violinist filled out two questionnaires, the \acrfull{crs}~\cite{knight2002comfort} and the \acrfull{stai}~\cite{speilbergermanual} scale. The complete questionnaires can be found in the supplementary materials.

\subsection{Data Description}
The raw data consisted of two main datasets: $\mathcal{D}=\langle D_{qnr}, D_{sns} \rangle$, where $D_{qnr}$ is a dataset consisting of all questionnaire responses, conforming to the tuple:
\begin{equation*}
   D_{qnr}=
\langle 
K = (user_{id}, trial_{index} \in (1 ... 29),
V
\rangle,
\end{equation*}
in which each data row is identified by a unique compound key $K$ being the combination of the participant identifier and the trial index designating any one of the excerpts that were played, and $V$ is a union of all attributes with questionnaire ratings on a $1...10$ scale. This includes raw features unified also with an attribute for the computed factor of \acrshort{eng} that was elicited as a latent construct as described in Section~\ref{sec: behave-model-dev}. 

The second dataset $D_{sns}$ is a time-series dataset conforming to the tuple:
\begin{equation*}
D_{sns}=
\langle
t,
K = (user_{id}, trial_{index} \in (1 ... 29),
S = (Sensor_{type}, M) 
\rangle, 
\end{equation*}
where $t$ is a timestamp of the measurement, $K$ is the key as in the first dataset, and $S$ is a sensor measurement denoting the type and its corresponding value. An exemplar dataset with data types, units, and descriptions is available at \url{github.com/AnonymousAuthors}.

\subsection{Behavioral Model Development for Data Labeling}
\label{sec: behave-model-dev}
\begin{figure}
  \begin{center}
    \includegraphics[width=0.6\linewidth]{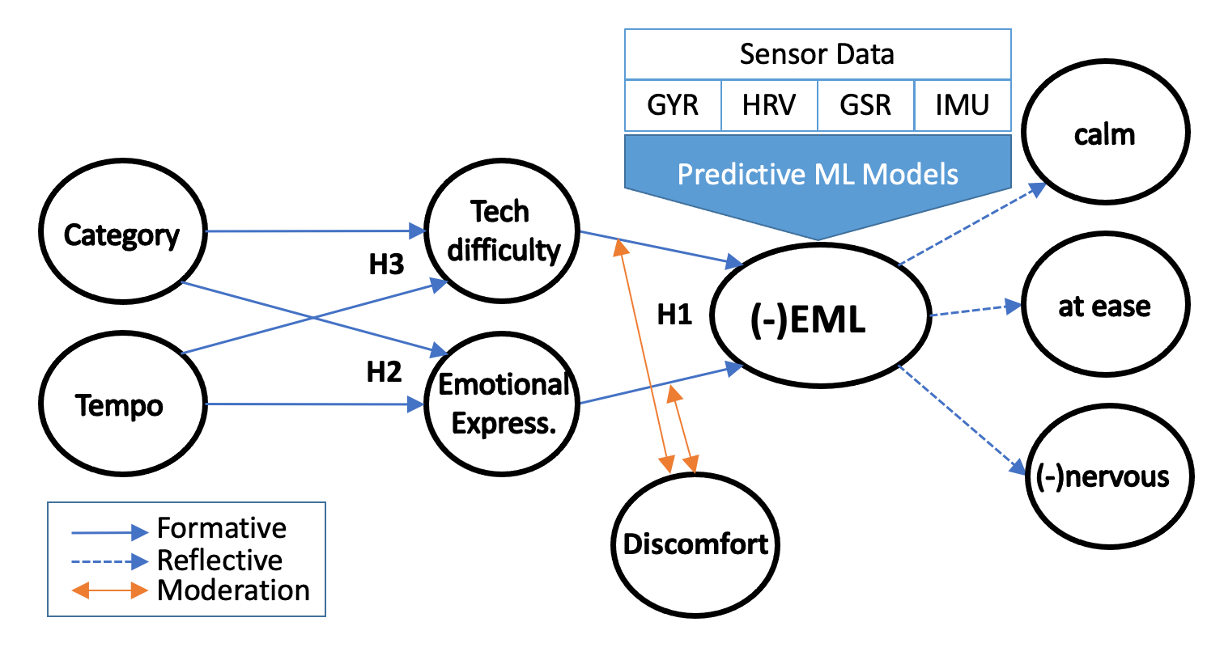}
  \end{center}
  \caption{Behavioral-model for the manipulation of \acrshort{eng}}
  \label{fig:engagement-mental-model}
\end{figure}
%Derived from `flow' literature, 

We began the gradual development of the behavioral model illustrated in Figure~\ref{fig:engagement-mental-model}. The objective was to establish a connection between the perceived manipulations of musical excerpts at different levels of \emph{technical difficulty} and experienced \emph{emotional expressiveness}, and the level of \acrshort{eng}. The purpose of this model development was twofold: (1) to ensure that our experiment is effectively affecting the level of \acrshort{eng}, and (2) to use the %evolving 
model for the labeling of our sensor-recordings with corresponding \acrshort{eng} levels. 
%as the target variable. % to then be predicted by the sensors. 
We also wanted to take into consideration the possible perception of discomfort that may be associated with wearing the sensors while playing the violin.
%This was designed to corroborate the following main experiment hypothesis:
Hence, our main experimental hypothesis was:

{\setlength{\parindent}{0cm}
\begin{hyp}[H\ref{hyp:first}] 
\label{hyp:first}
\ul{Technical difficulty} and \ul{emotional expressiveness} for a played musical excerpt, as perceived surrogates of the manipulation, affect the learner's \underline{engagement} in the excerpt-playing task.
\end{hyp}}

% \begin{figure}
% \centering
% \includegraphics[width=1.0\textwidth]{figures/engagement-model-new.png}
% \caption{behavioral-model for the manipulation of \acrshort{eng}}
% \label{fig:engagement-mental-model}
% \end{figure}

In order to effectively influence the perceptions of technical difficulty and emotional expressiveness, we employed a combination of two factors: excerpt \emph{tempo} and \emph{type} (i.e., musical category) as our direct stimuli. Overall, the collection of the excerpts assigned to each musician included excerpts of two types: 24 \emph{technical} and 5 \emph{repertoire}. For the technical type, the musicians were instructed to play 12 excerpts at a \emph{slow} tempo, and 12 at a \emph{fast} tempo (randomly ordered in the experimental protocol). For the repertoire type, musicians were instructed to play one at a \emph{fast} tempo, two at an \emph{average} tempo, and two at a \emph{slow} tempo. This resulted in each musician playing a total of 29 sessions. As aforementioned, after each excerpt, the musician reported (on a 1--10 scale) their perceptions of technical-difficulty and of emotional-expressiveness. With respect to this setup, we hypothesized that:

{\setlength{\parindent}{0cm}
\begin{hyp}[H\ref{hyp:second}] 
\label{hyp:second}
Excerpt's \ul{musical-category} and played \ul{tempo} affect learner's perceived \ul{emotional-expressiveness}.
\end{hyp}}

% \begin{hyp}[H\ref{hyp:second}] 
% \label{hyp:second}
% Excerpt's musical-category and played tempo affect learner's perceived emotional-expressiveness.
% \end{hyp}

{\setlength{\parindent}{0cm}
\begin{hyp}[H\ref{hyp:third}] 
\label{hyp:third}
Excerpt's \underline{musical-category} and played \underline{tempo} affect learner's perceived \underline{technical-difficulty}.
\end{hyp}}

The data collected in the questionnaires were examined to test for these two preceding desired effects. We conducted two-way ANOVA analyses to test for the possible effects in H\ref{hyp:second} and H\ref{hyp:third}.
% \footnote{The mean and STD of the measures tech\_diff, emo\_expr, feel\_calm, feel\_at\_ease, feel\_nervous and feel\_uncomfortable are $(4.12, 2.405)$, $(4.36, 2.699)$, $(7.03, 2.089)$, $(7.11, 2.041)$, $(3.20, 2.069)$,and $(4.07, 1.961)$ respectively.}

\begin{table}[ht]
\centering
\caption{Descriptive statistics for all measures}
\label{tab:measures-statisticfs}

\begin{tabular}{|l|l|l|l|l|l|}
\hline
\textbf{Measure} & \textbf{N} & \textbf{Min} & \textbf{Max} & \textbf{Mean} & \textbf{Std. Dev} \\ \hline
tech\_diff          & 261 & 1 & 10 & 4.12 & 2.405 \\ \hline
emo\_expr           & 261 & 1 & 10 & 4.36 & 2.699 \\ \hline
feel\_calm          & 261 & 1 & 10 & 7.03 & 2.089 \\ \hline
feel\_at\_ease      & 261 & 1 & 10 & 7.11 & 2.041 \\ \hline
feel\_nervous       & 261 & 1 & 10 & 3.20 & 2.069 \\ \hline
feel\_uncomfortable & 261 & 1 & 10 & 4.07 & 1.961 \\ \hline
\end{tabular}

\end{table}

With respect to the effect on emotional-expressiveness (H\ref{hyp:second}), the two-way interaction musical-category$*$tempo was not found to be significant, where only the main effect of musical-category was significant ($F_{(1,256)}=177.239$)\footnote{p-value was smaller than .001 if not reported otherwise}. There was no significant effect of the tempo at which the excerpt is played on emotional-expressiveness.

% With respect to H\ref{hyp:third}, both main effects of category ($F_{(1,256)}=19.999, p<.001$) and tempo ($F_{(2,256)}=50.611, p<.001$) were found to be significant, but also the two-way interaction between musical-category$*$tempo was found to be significant ($F_{1,256}=13.310,p<.001$). Post-hoc pair-wise comparisons %(see Figure~\ref{fig:post-hocs})
% indicated significant differences in the perception of technical-difficulty between the categories when the tempo is fast ($F_{1,115}=23.611,p<.001
% $). Average tempo was considered only in the repertoire type. There was no significant difference in the perception of technical-difficulty between the categories when the tempo is slow. 

With respect to H\ref{hyp:third}, both main effects of the category ($F_{(1,256)}=19.999 $) and tempo ($F_{(2,256)}=50.611$) were found to be significant, but also was the two-way interaction between musical-category$*$tempo %was found to be significant 
($F_{1,256}=13.310$). Thus, post-hoc pair-wise comparisons %(see Figure~\ref{fig:post-hocs})
indicated significant differences in the perception of technical-difficulty between the categories when the tempo is fast ($F_{1,115}=23.611
$). The average tempo was considered only in the repertoire type. There was no significant difference in the perception of technical-difficulty between the categories when the tempo is slow. 

% \begin{figure}[ht]
%     \begin{center}
%     % \centering
%     \includegraphics[width=0.6\linewidth]{fig_spss/post_hoc_tech_diff_tempo.png}
%     \end{center}
%     \caption{post-hoc category vs. tempo}
%     \label{fig:post-hocs}
% \end{figure}

Hence, the sole effect of the category manipulation is inconsistent across different levels of tempo, such that for a slow tempo, there is not much of a difference in the perception of technical-difficulty when the type of excerpt is altered. Therefore, it was essential to combine the manipulation of the excerpt-category jointly with the manipulation of tempo to ensure the perception of technical-difficulty is altered. Nevertheless, we concluded that our combined experimental stimuli were effective in altering both perceptions of technical-difficulty and emotional-expressiveness, allowing us to continue with the examination of the subsequent effect on \acrshort{eng} as hypothesized in H\ref{hyp:first}.

For labeling, we aimed to capture the extent to which each musician was immersed in the act of playing the excerpt. For this, we attended to the literature notion of being engaged in a task or being in a `flow', %state 
as articulated by the Flow-Theory~\cite{csikszentmihalyi1990flow}. In line with this view, optimal \acrshort{eng} is attained when in a proper balance between one's level of personal competence and the perceived complexity level of the task at hand. As illustrated in Figure~\ref{fig:flow-theory}, the flow condition along the diagonal is reflected by a sweat-spot between anxiety and boredom. Particularly, the inclusion of more competent musicians in our experiment as corroborated by the \acrshort{gmsi} index, let us focus on the right-hand side of this chart, where \acrshort{eng} may be conceived as balancing nervousness and relaxation along an increase in complexity of the task. To capture such a condition, %at the end of 
after playing each excerpt, we employed %key measures for 
self-reporting of calmness, being at ease, and nervousness perceptions using measures extracted from the \acrshort{stai} scale that was fully administered at the end of the experiment.  

While \acrshort{eng} is formed during the actual playing of a musical excerpt and manipulated via the aforementioned characteristics of the excerpts (i.e., technical-difficulty and emotional-expressiveness), the three measures of calmness, being at ease, and nervousness are expected to be mutually \emph{reflective}~\cite{formative-reflective-2007} of the level of \acrshort{eng} as a latent construct. These measurements require conventional validation through Cronbach's Alpha (reliability) and Factor Analysis (convergence and discriminant validity). Respectively, reliability among the three measures was verified with Cronbach's $\alpha=.922$. Principal component analysis was conducted, considering Eigenvalues greater than 1 (see Table~\ref{tab:factor-analysis}). The minimum factor loading criterion was set to 0.50. The communality of the scale, indicating the amount of variance in each dimension, was also assessed to ensure acceptable levels of explanation. The results of this analysis confirmed a single component (namely \acrshort{eng}) for all three measures. The factor loadings and communalities for the three factors, feel\_calm, feel\_at\_ease, and inv\_nervous, are (0.972, 0.945), (0.931, 0.867), and (0.784, 0.614), respectively.

\begin{table}[ht]

\centering
\caption{\textbf{Factor Analysis:\\} {feel\_calm, inv\_nervous, feel\_at\_ease $\xrightarrow{}$ nirvana}}
\label{tab:factor-analysis}
\begin{tabular}{l l l l l l}
\hline
\multicolumn{6}{c}{Total Variance Explained}\\ \hline
& \multicolumn{3}{l}{Initial Eigenvalues} & \multicolumn{2}{l}{Sums of Squared Loadings} \\
{Factor}  & {Total}  & {\% of Variance} & {Cumulative \%} & {Total}  & {\% of Variance}  \\ \hline
1  & 2.600  & 86.662  & 86.662  & 2.425  & 80.843   \\ \hline
2  & 0.307  & 10.234  & 96.896  &   &    \\ \hline
3  & 0.093  & 3.104  & 100.000  &   &    \\ \hline
\end{tabular}
% }
\end{table}

Following dimensionality reduction %as described 
for \acrshort{eng} as the latent construct, we then moved on to testing our main %experimental 
hypothesis (i.e., H\ref{hyp:first}). %That is, to test for the effects of technical-difficulty and emotional-expressiveness on the latent reflective factor of \acrshort{eng}. 
Additionally, we took into consideration the potential effect %that may occur as a result of the level 
of perceived discomfort associated with wearing additional sensors while playing the violin. Therefore, we speculated that the perception of discomfort might moderate the hypothesized main effects %we 
%hypothesized %about 
(i.e., a high sense of discomfort could potentially hinder the experienced \acrshort{eng}). To explore this, we conducted a two-way ANCOVA analysis for H\ref{hyp:first}, with discomfort included as a covariate. The results, as presented in Table~\ref{tab:ancova-for-h1}, indicate significant main effects on \acrshort{eng} for both technical-difficulty ($F_{(9,185)}=6.492,p<.001$) and emotional-expressiveness ($F_{(9,185)}=50.584,p=.002$). Moreover, the covariate, discomfort was found to be significantly associated with \acrshort{eng} ($F_{(1,185)}=7.59,p=.006$).

\begin{table}[ht]

\centering
\caption{\textbf{Ancova for H\ref{hyp:first}:} tech\_diff, emot\_expr (independent), discomfort (covariant) $->$ \acrshort{eng}}
\label{tab:ancova-for-h1}
% \caption{Factor Analysis: Feel calm, inverse nervous, feel at_ease to Engagement}
% \small
% \resizebox{\columnwidth}{!}{
\begin{tabular}{l l l l l l}
\hline
\multicolumn{6}{c}{Tests of Between-Subjects Effects}\\ \hline
% & \multicolumn{3}{l}{Initial Eigenvalues} & \multicolumn{3}{l}{Extraction Sums of Squared Loadings} \\
{Source}  & {Type 3 SSE}  & {df} & {Mean Square} & {F}  & {Sig.}  \\ \hline
Corrected Model&        4171.661$^a$&   75 & 55.622     & 3.379  & $<$0.001 \\ \hline
Intercept&              8210.182&       1  & 8210.182   & 498.697 & $<$0.001\\ \hline
feel\_uncomfortable&    124.962&        1  & 124.962    & 7.590 & 0.006 \\ \hline
emo\_expr&              455.260&        9  & 50.584     & 3.073 & 0.002  \\ \hline
tech\_diff&             961.923&        9  & 106.880    & 6.492 & $<$0.001 \\ \hline
emo\_expr * tech\_diff& 1269.709&       56  & 22.673    & 1.377 & 0.059  \\ \hline
Error&                  3045.704&       185  & 16.463   &       &   \\ \hline
Total&                  102481.211&     261  &          &       &    \\ \hline
Corrected Total&        7217.365&       260  &          &       &    \\ \hline
    \end{tabular}
    % }
\end{table}

The effect size of the statistical analyses is as follows:
Technical difficulty  13.3\%, Emotional expressiveness 6.3\%, Discomfort 1.7\%, Interaction (tech * emotional exp) 17.5\% (not significant), Error 42.1\%, and total for model ($R^2$) 57.8\%.

Concluding the effort in developing the behavioral model, we have successfully created and validated the necessary instrumentation for manipulating and measuring \acrshort{eng} through self-reporting. We have also identified that the perception of discomfort may have a moderating effect on \acrshort{eng}. Hence, we used the developed model to label the sensor dataset, where we partitioned between `high' and `low' \acrshort{eng} as was naturally split by its mean value. %In addition, 
Additionally, we utilized the reporting of discomfort as another target label. This was done to allow for the development of two \acrshort{ml} models with the acquired sensor data: 
% (1) prediction of discomfort, and (2) classification of \acrshort{eng} utilizing the output from model (1) as an additional input feature, given its significant moderation effect.
(1) discomfort prediction, and (2) classification of \acrshort{eng}. The output from model (1) was used as an additional input feature for model (2) due to its significant moderating effect.

\section{Stage II: Machine learning model development}\label{sec:ml}

The primary objective of this stage was to develop a \acrshort{ml} model that can classify a musician's \acrshort{eng} while playing the violin, utilizing the sensor data that were labeled in the previous stage.

\subsection{Data preparation for ML}\label{Data preparation for ML}
We established two parameters, namely \textit{step-size} and \textit{window-size}, and employed them to partition each excerpt's sensor dataset into windows. For example, if the \textit{step-size} is set to $20$ seconds and the \textit{window-size} is set to $30$ seconds, the dataset for each excerpt is divided into a sequence of $30$-second windows with a $10$-second overlap between any two adjacent windows. The motivation for such windowing is two-fold. Firstly, it allows us to be more responsive in terms of the time required for making a prediction. If we do not partition the excerpt, predictions can only be made per expert, which occurs approximately every $90$ seconds on average. Secondly, for the application of machine learning algorithms, a sufficient sample size is needed ($261$ data points without windowing). The concrete determination of these parameters is discussed in Section~\ref{sec:results}. %the Results section, 
%following rigorous sensitivity analysis.

%In each window, 
A total of $257$ features were computed from the sensor data for each window. We used out-of-box sensor libraries such as PyEDA and PyHRV to extract the features. The sensor data were enriched with a first derivative termed ``jerk'', defined as the difference in values divided by the difference in time between two adjacent entries. It was computed for the \acrshort{imu} and \acrshort{gyr} data. In addition, the magnitude of the 3-axis sensors was determined by taking the square root of the sum of squares of the sensor values. The magnitude calculation was performed for both the \acrshort{imu} and \acrshort{gyr} data, along with their respective jerk values, across the x, y, and z axes.

From the \acrshort{imu} and \acrshort{gsr} sensors, %(two positions)
we extracted several features including the interquartile ratio, kurtosis, mean, median, RMSSD (the root-mean-square of differences between successive RR intervals), skewness, and variance of both magnitude and jerk-magnitude. Specifically, from the \acrshort{imu} sensor, we also extracted the SPARC index~\cite{sparc-index} as a feature, which denotes the smoothness of movement associated with the \acrshort{imu} positioned on the forearm.
From the \acrshort{hrv} sensor, %(one position) 
we extracted various features including the mean, median, frequency domain features (e.g., LF, HF, VLF), %, total power), 
time domain features (e.g., RMSSD, NNI@20, NNI@50, PNNI@20, PNNI@50, NNI range, SDNN), as well as non-linear measures (e.g., SampEntropy, SD1, SD1/SD2).
From the \acrshort{gsr} sensors, %(3 positions) 
features were extracted with the PyEDA library, catering to a combination of statistical features such as the number of peaks, amplitude mean, variance, and automatic features using autoencoding.

Prior to feature extraction, we investigated the correspondence between the \acrshort{gsr} and \acrshort{imu} sensors, to accommodate for any undesired movement artifacts~\cite{gsr-movement-2012} that could interfere with the accuracy of the \acrshort{gsr} sensor readings. Specifically, we performed a Pearson correlation test to examine the correlation between the readings of each individual \acrshort{gsr} sensor and the \acrshort{imu}. The results indicated that both correlations were insignificant. Therefore, there was no need to discard any of the \acrshort{gsr} measurements, as they were not affected by the movement artifacts captured by the \acrshort{imu} sensor.

Post-feature extraction, we used the eye-closed task (baseline sessions) for inter-subject normalization to account for the individual differences among participants. For each $\langle$user X feature$\rangle$ in the baseline trials, a corresponding mean value was calculated. Subsequently, each feature of the non-baseline excerpts was divided by the corresponding baseline mean value of that feature.

Each window was labeled with the \acrshort{eng} level (`high' or `low') and discomfort level, both based on the scores  determined by the mental model that was developed in the previous stage. It is important to note that the original data labels were determined at the level of each excerpt as a whole, based on the original musician's indications reported at the end of each excerpt. We then interpolated each value across all windows that correspond to the same excerpt. This hints at an assumption that the user's \acrshort{eng} and sense of discomfort are two relatively stable conditions that do not undergo significant changes during the course of playing an individual excerpt.

\subsection{Predicting the discomfort level}
\begin{figure}[h!]
\label{eng_predic_pipline_discom}
    \centering
    \resizebox{0.9\linewidth}{!}{
\tikzset{every picture/.style={line width=0.9pt}} %set default line width to 0.75pt        

\begin{tikzpicture}[x=0.75pt,y=0.75pt,yscale=-1,xscale=1]
%uncomment if require: \path (0,231); %set diagram left start at 0, and has height of 231

%Rounded Rect [id:dp21299074500456994] 
\draw   (116,24.8) .. controls (116,16.63) and (122.63,10) .. (130.8,10) -- (175.2,10) .. controls (183.37,10) and (190,16.63) .. (190,24.8) -- (190,98.2) .. controls (190,106.37) and (183.37,113) .. (175.2,113) -- (130.8,113) .. controls (122.63,113) and (116,106.37) .. (116,98.2) -- cycle ;
%Straight Lines [id:da7395819321049393] 
\draw    (64,99) -- (114,99) ;
\draw [shift={(116,99)}, rotate = 180] [color={rgb, 255:red, 0; green, 0; blue, 0 }  ][line width=0.75]    (10.93,-3.29) .. controls (6.95,-1.4) and (3.31,-0.3) .. (0,0) .. controls (3.31,0.3) and (6.95,1.4) .. (10.93,3.29)   ;
%Straight Lines [id:da19355084286230584] 
\draw    (64,24) -- (114,24) ;
\draw [shift={(116,24)}, rotate = 180] [color={rgb, 255:red, 0; green, 0; blue, 0 }  ][line width=0.75]    (10.93,-3.29) .. controls (6.95,-1.4) and (3.31,-0.3) .. (0,0) .. controls (3.31,0.3) and (6.95,1.4) .. (10.93,3.29)   ;
%Straight Lines [id:da4280670756062025] 
\draw    (64,80.25) -- (114,80.25) ;
\draw [shift={(116,80.25)}, rotate = 180] [color={rgb, 255:red, 0; green, 0; blue, 0 }  ][line width=0.75]    (10.93,-3.29) .. controls (6.95,-1.4) and (3.31,-0.3) .. (0,0) .. controls (3.31,0.3) and (6.95,1.4) .. (10.93,3.29)   ;
%Straight Lines [id:da01644034709378328] 
\draw    (64,61.5) -- (114,61.5) ;
\draw [shift={(116,61.5)}, rotate = 180] [color={rgb, 255:red, 0; green, 0; blue, 0 }  ][line width=0.75]    (10.93,-3.29) .. controls (6.95,-1.4) and (3.31,-0.3) .. (0,0) .. controls (3.31,0.3) and (6.95,1.4) .. (10.93,3.29)   ;
%Straight Lines [id:da6784973043185042] 
\draw    (64,42.75) -- (114,42.75) ;
\draw [shift={(116,42.75)}, rotate = 180] [color={rgb, 255:red, 0; green, 0; blue, 0 }  ][line width=0.75]    (10.93,-3.29) .. controls (6.95,-1.4) and (3.31,-0.3) .. (0,0) .. controls (3.31,0.3) and (6.95,1.4) .. (10.93,3.29)   ;
%Straight Lines [id:da7945294531024377] 
\draw    (191,99) -- (241,99) ;
\draw [shift={(243,99)}, rotate = 180] [color={rgb, 255:red, 0; green, 0; blue, 0 }  ][line width=0.75]    (10.93,-3.29) .. controls (6.95,-1.4) and (3.31,-0.3) .. (0,0) .. controls (3.31,0.3) and (6.95,1.4) .. (10.93,3.29)   ;
%Straight Lines [id:da20001104063754593] 
\draw    (191,24) -- (241,24) ;
\draw [shift={(243,24)}, rotate = 180] [color={rgb, 255:red, 0; green, 0; blue, 0 }  ][line width=0.75]    (10.93,-3.29) .. controls (6.95,-1.4) and (3.31,-0.3) .. (0,0) .. controls (3.31,0.3) and (6.95,1.4) .. (10.93,3.29)   ;
%Straight Lines [id:da7824119692722447] 
\draw    (191,80.25) -- (241,80.25) ;
\draw [shift={(243,80.25)}, rotate = 180] [color={rgb, 255:red, 0; green, 0; blue, 0 }  ][line width=0.75]    (10.93,-3.29) .. controls (6.95,-1.4) and (3.31,-0.3) .. (0,0) .. controls (3.31,0.3) and (6.95,1.4) .. (10.93,3.29)   ;
%Straight Lines [id:da8042508050252782] 
\draw    (191,61.5) -- (241,61.5) ;
\draw [shift={(243,61.5)}, rotate = 180] [color={rgb, 255:red, 0; green, 0; blue, 0 }  ][line width=0.75]    (10.93,-3.29) .. controls (6.95,-1.4) and (3.31,-0.3) .. (0,0) .. controls (3.31,0.3) and (6.95,1.4) .. (10.93,3.29)   ;
%Straight Lines [id:da3308959932479949] 
\draw    (191,42.75) -- (241,42.75) ;
\draw [shift={(243,42.75)}, rotate = 180] [color={rgb, 255:red, 0; green, 0; blue, 0 }  ][line width=0.75]    (10.93,-3.29) .. controls (6.95,-1.4) and (3.31,-0.3) .. (0,0) .. controls (3.31,0.3) and (6.95,1.4) .. (10.93,3.29)   ;
%Rounded Rect [id:dp20548466314882974] 
\draw   (243,25.6) .. controls (243,17.43) and (249.63,10.8) .. (257.8,10.8) -- (302.2,10.8) .. controls (310.37,10.8) and (317,17.43) .. (317,25.6) -- (317,99) .. controls (317,107.17) and (310.37,113.8) .. (302.2,113.8) -- (257.8,113.8) .. controls (249.63,113.8) and (243,107.17) .. (243,99) -- cycle ;
%Curve Lines [id:da6418530510712965] 
\draw    (217,42.75) .. controls (215,165.75) and (214,149.75) .. (352,150.35) ;
\draw [shift={(352,150.35)}, rotate = 180.25] [color={rgb, 255:red, 0; green, 0; blue, 0 }  ][line width=0.75]    (10.93,-3.29) .. controls (6.95,-1.4) and (3.31,-0.3) .. (0,0) .. controls (3.31,0.3) and (6.95,1.4) .. (10.93,3.29)   ;
%Rounded Rect [id:dp6486726369098613] 
\draw   (352,112.6) .. controls (352,102.88) and (359.88,95) .. (369.6,95) -- (422.4,95) .. controls (432.12,95) and (440,102.88) .. (440,112.6) -- (440,202.2) .. controls (440,211.92) and (432.12,219.8) .. (422.4,219.8) -- (369.6,219.8) .. controls (359.88,219.8) and (352,211.92) .. (352,202.2) -- cycle ;
%Curve Lines [id:da5545473710144997] 
\draw    (211,62) .. controls (209,185) and (214,168.4) .. (352,169) ;
\draw [shift={(352,169)}, rotate = 180.25] [color={rgb, 255:red, 0; green, 0; blue, 0 }  ][line width=0.75]    (10.93,-3.29) .. controls (6.95,-1.4) and (3.31,-0.3) .. (0,0) .. controls (3.31,0.3) and (6.95,1.4) .. (10.93,3.29)   ;
%Curve Lines [id:da41785951251706943] 
\draw    (204,81) .. controls (202,204) and (213,186.4) .. (351,187) ;
\draw [shift={(351,187)}, rotate = 180.25] [color={rgb, 255:red, 0; green, 0; blue, 0 }  ][line width=0.75]    (10.93,-3.29) .. controls (6.95,-1.4) and (3.31,-0.3) .. (0,0) .. controls (3.31,0.3) and (6.95,1.4) .. (10.93,3.29)   ;
%Curve Lines [id:da3684251131235262] 
\draw    (198,99) .. controls (196,222) and (214,204.4) .. (352,205) ;
\draw [shift={(352,205)}, rotate = 180.25] [color={rgb, 255:red, 0; green, 0; blue, 0 }  ][line width=0.75]    (10.93,-3.29) .. controls (6.95,-1.4) and (3.31,-0.3) .. (0,0) .. controls (3.31,0.3) and (6.95,1.4) .. (10.93,3.29)   ;
%Curve Lines [id:da5516193080492896] 
\draw    (224,24) .. controls (222,147) and (214,131) .. (352,131.6) ;
\draw [shift={(352,131.6)}, rotate = 180.25] [color={rgb, 255:red, 0; green, 0; blue, 0 }  ][line width=0.75]    (10.93,-3.29) .. controls (6.95,-1.4) and (3.31,-0.3) .. (0,0) .. controls (3.31,0.3) and (6.95,1.4) .. (10.93,3.29)   ;
%Curve Lines [id:da2080171214764559] 
\draw    (317,59) .. controls (346.7,61.97) and (302.89,115.91) .. (350.53,115.04) ;
\draw [shift={(352,115)}, rotate = 177.71] [color={rgb, 255:red, 0; green, 0; blue, 0 }  ][line width=0.75]    (10.93,-3.29) .. controls (6.95,-1.4) and (3.31,-0.3) .. (0,0) .. controls (3.31,0.3) and (6.95,1.4) .. (10.93,3.29)   ;

%Straight Lines [id:da01734908720942896] 
\draw    (440,159) -- (474,159.47) ;
\draw [shift={(476,159.5)}, rotate = 180.8] [color={rgb, 255:red, 0; green, 0; blue, 0 }  ][line width=0.75]    (10.93,-3.29) .. controls (6.95,-1.4) and (3.31,-0.3) .. (0,0) .. controls (3.31,0.3) and (6.95,1.4) .. (10.93,3.29)   ;
%Shape: Pulse Wave Form [id:dp22052104704643882] 
\draw   (479,169) -- (501.22,169) -- (501.22,148.5) -- (536.78,148.5) -- (536.78,169) -- (559,169) ;

% Text Node
\draw (119,36) node [anchor=north west][inner sep=0.75pt]  [font=\large] [align=left] {Windowing\\$\displaystyle \&\ $Feature\\Extraction};
% Text Node
\draw (15,34) node [anchor=north west][inner sep=0.75pt]  [font=\large] [align=left] {Raw\\Sensor\\Data};
% Text Node
\draw (247,53) node [anchor=north west][inner sep=0.75pt]  [font=\large] [align=left] {Predict\\Discomfort};
% Text Node
\draw (357,142) node [anchor=north west][inner sep=0.75pt]  [font=\large] [align=left] {Predict\\Engagement\\in Motor\\Learning};
% Text Node
\draw (565,159) node [anchor=north west][inner sep=0.75pt]   [font=\large][align=left] {Low EML};
% Text Node
\draw (541,138) node [anchor=north west][inner sep=0.75pt]   [font=\large][align=left] {High EML};

\end{tikzpicture}
}
    \caption{ML engagement prediction pipeline}
    \label{fig:ml-pipeline}
\end{figure}
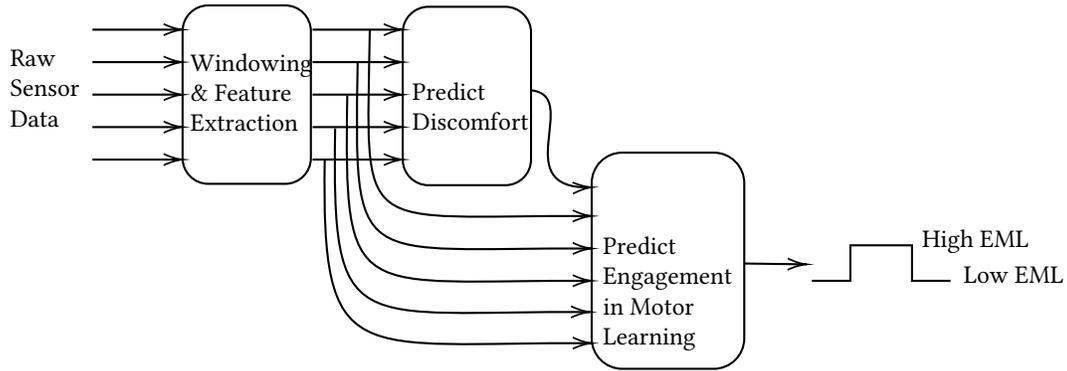

During the development of the mental model, we observed that the discomfort level is a covariant with respect to the \acrshort{eng}. We thus presume that an \acrshort{eng} classification model may benefit if it also utilizes the discomfort level as an input. This has required us to develop a designated model for the prediction of discomfort with sensor data as an input (see Figure~\ref{fig:ml-pipeline}).

\begin{figure}
    \centering
    \includegraphics[width=0.6\textwidth]{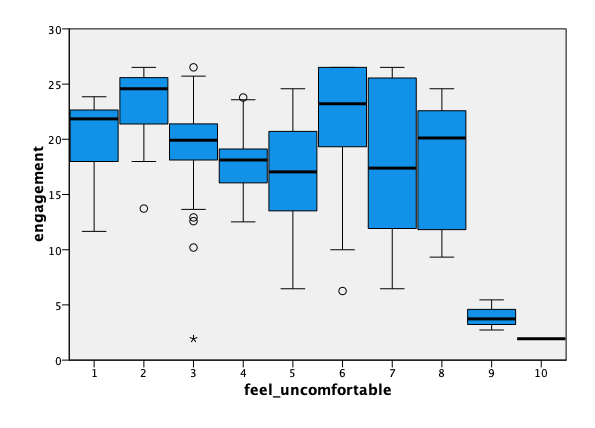}
    \caption{Box-plot of \acrshort{eng} (`engagement') levels for each discomfort level}
    \label{fig:discomfort_bins}
\end{figure}

As an initial step, we examined the relationship between feeling discomfort and \acrshort{eng} using box-plot distributions, as shown in Figure~\ref{fig:discomfort_bins}.
% This chart revealed the most fundamental difference occurring between the discomfort segments of $[1-8]$ and $[9-10]$, where \acrshort{eng} drops extensively in the latter.
This chart highlighted the most profound difference between the discomfort segments of $[1-8]$ and $[9-10]$, with \acrshort{eng} dropping significantly in the latter.
Accordingly, we split the data into bins labeled `high discomfort' and `normal discomfort'.

Utilizing the discomfort labeling, we trained a binary classification model that takes the sensor data as input and predicts whether the discomfort level is `normal' or `high'. We performed a $5$-fold cross-validation \cite{kohavi1995study}. An average accuracy level of $0.98$ was achieved with an SVM classifier without hyperparameter optimization. The two most important features for classification were found to be the variance of the \acrshort{imu} and the variance of the \acrshort{gyr}.

This suggests that we can easily track the discomfort level with the sensors, and we do not need a sophisticated ML model to do so.
This should not come as a surprise, as the discomfort level is highly correlated to the variance of the \acrshort{gyr} sensor.

% \lior{edited with https://www.mathcha.io/editor}

\section{Results}
\label{sec:results}

% In terms of the data, we randomly partitioned it into 5 parts, with each part serving as a test set once, while the remaining 4 parts were used for training
We partitioned the data randomly into five parts. Each part served as a test set once, while the other four parts were used for training (i.e., 5-fold cross-validation with $20\%$ for testing and $80\%$ for training \cite{kohavi1995study}). The partitioning was carefully designed to prevent data leakage between training and test sets. That is, during the preparation of the data for the five-fold cross-validation, we ensured that data from the same parts of the excerpts did not exist in both the training and testing set.

\begin{figure}[h!]
    \centering
    \includegraphics[width=0.8\linewidth]{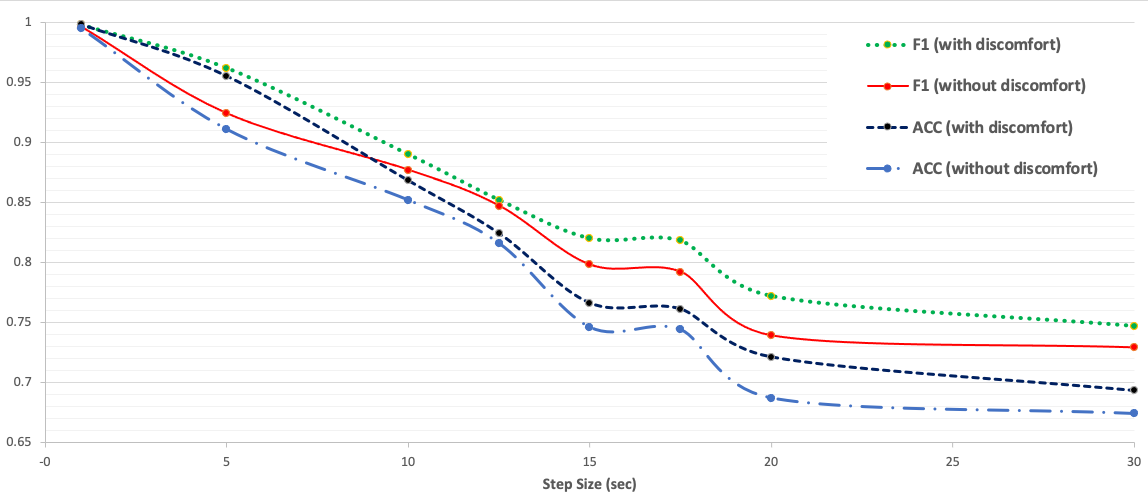}
    \caption{$F_1$ and $ACC$ scores with $5$-fold cross validation for different \textit{step-size} values; with \textit{window-size}  of $30$ seconds.}
    \label{fig:ml_results}
\end{figure}

We report the average results for a fixed $30$-second \textit{window-size} with the top-performing XGBoost algorithm in Figure \ref{fig:ml_results}. Both $F_1$ and $accuracy$ scores were observed. The class balance was $54\%$ for `low' \acrshort{eng} and $46\%$ for `high' \acrshort{eng}.

% We also performed a sensitivity check both for the window size and for the step size with the top-performing XGBoost algorithm. For window size, we followed a convention suggested in \cite{zeltyn2016enhanced} on general activity recognition and fixed a $30$-second \textit{window-size}. Manipulating around it was only marginally affecting the results. Hence, a \textit{window-size} of $30$ seconds was chosen as a sweet spot. It was long enough to provide sufficient informative sensor data for model training, yet short enough %so that enough data points will be created, while 
% to maintain model responsiveness by providing predictions every few seconds. 

We conducted a sensitivity check for both the \textit{window size} and the \textit{step size} using the top-performing XGBoost algorithm. Regarding the window size, we followed the convention suggested in \cite{zeltyn2016enhanced} for general activity recognition and set it to a fixed duration of $30$ seconds. Modifying it only had a marginal impact on the results. Therefore, a \textit{window size} of $30$ seconds was chosen as a sweet spot. This duration was long enough to provide sufficient informative sensor data for model training, yet short enough to maintain model responsiveness by generating predictions every few seconds.

With respect to the manipulation of step size, as illustrated in Figure \ref{fig:ml_results}, we decided to set it at $15$ seconds. This means that the model would provide an initial output after 30 seconds and subsequent predictions every 15 seconds. 
% The overlapping windows ensured a certain degree of statefulness in the evolution of \acrshort{eng}, and we acknowledged the possibility of employing a model like LSTM, which inherently incorporates a `memory' component in its architecture.
The overlapping windows ensured a degree of statefulness in the evolution of \acrshort{eng}. We recognized the potential of using a model like LSTM, which inherently has a 'memory' component in its architecture.
We interpreted the increase in accuracy for a shorter \textit{step-size} (below the bending point at 15 seconds) as mainly stemming from a synthetic improvement resulting from extensive window overlapping, where more than half of a window overlaps with the adjacent windows. We also tested with different step sizes (e.g., 30) and observed that most results remained consistent.

For the construction of the main \acrshort{ml} model for \acrshort{eng}, 
% we used a commercial AI service that provided us with access to a wide range of models.
we resorted to a commercial AI service, which had a variety of models available.
The key models we experimented with included Decision Tree, Gradient Boosting, LGBM, Logistic Regression, Random Forest, SVM Classifier, and XGBoost.
We opted to use the default hyperparameters for all of these algorithms. Although the commercial tool offered automatic hyperparameter optimization, the resulting improvements were found to be insignificant.

The top-performing result was achieved by the XGBoost classifier, which obtained an $F_1$ score of $0.82$ when the discomfort feature was included in its input. When the discomfort level was ignored, the performance of the models decreased by approximately 5\%. We also conducted tests comparing the XGBoost classifier with other algorithms. We employed a non-parametric bootstrapping test  \cite{efron1994introduction,dror-etal-2019-deep} and used 5-fold cross-validation for the evaluation of the results. The XGBoost classifier demonstrated a significant improvement ($p<0.01$) compared to the other algorithms when the \textit{step-size} was larger than $10$. As shown in Table \ref{tab:conf_matrix}, the classifier demonstrates good results for both the ``low'' and ``high'' engagement classes.

% \begin{table}[ht]
%         \caption{A confusion matrix for one of the 5-fold  validation.}
%     \label{tab:conf_matrix}
%     \centering
% \resizebox{0.8\linewidth}{!}{
% \begin{tabular}{c|c c |c}
%     \hline
%     &  \multicolumn{3}{c||}{Predicted \acrshort{eng} }   \\
%          Observed \acrshort{eng}    & `high' & `low' & Percent correct   \\
%          \hline\hline
%          `high'  &  \textbf{41} & 8 & 83.7\% \\ 
%          `low' & 7 & \textbf{62} & 89.9\% \\ 
%          \hline
%          Percent correct & 85.4\% & 88.6\% & \textbf{87.3}\% \\
%          \hline
%     \end{tabular}}
% \end{table}

\begin{table}[ht]
        \caption{A confusion matrix for one of the 5-fold  validation.}
    \label{tab:conf_matrix}
    \centering
\begin{tabular}{cc c c}
    \hline
    &       \multicolumn{2}{c}{Predicted \acrshort{eng}}      \\
    Observed \acrshort{eng} & `high' & `low' & Percent correct   \\
         \hline
         `high'  &  \textbf{41} & 8 & 83.7\% \\ 
         `low' & 7 & \textbf{62} & 89.9\% \\ 
         Percent correct & 85.4\% & 88.6\% & \textbf{87.3}\% \\
         \hline
    \end{tabular}
\end{table}

% As can be seen in Table \ref{tab:conf_matrix}, the classifier achieves good results on both the ``low'' and ``high'' engagement classes.

In addition to the XGBoost classifier, the three top-performing algorithms were the LGBM classifier, the Decision Tree Classifier, and logistic regression. Table \ref{tbl:Feature_importance} presents the feature importance for these different algorithms. For tree-based classification algorithms like Decision Tree, XGBoost, and LGBM, the feature importance represents their inherent importance scores, calculated based on the reduction in the criterion used to select split points during training. For non-tree algorithms such as Logistic Regression, the feature importance reflects the importance determined by training a Random Forest algorithm on the same training data as the non-tree algorithm. For each sensor type, the most important feature is displayed.

\begin{table}[ht]
\caption{Feature importance.}
\label{tbl:Feature_importance}
\begin{tabular}{ccccc}
\hline\
                    & \acrshort{imu}                 & \acrshort{gyr}                & \acrshort{gsr}        & \acrshort{hrv}       \\ \hline
XGBoost             & elbow\_jerk\_var     & elbow\_rmssd        & left\_axilla & high\_frequency        \\
LGBM                & elbow\_kurt          & wrist\_jerk\_kurt   & left\_axilla & mean\_nni \\
Decision Tree       & wrist\_magnitude\_rmssd & elbow\_jerk\_median & left\_axilla & high\_frequency        \\
Logistic Regression & wrist\_magnitude\_rmssd & elbow\_jerk\_median & left\_axilla & high\_frequency       \\\hline
\end{tabular}
\end{table}
\vspace{0.2cm}
\begin{table}[ht]
\caption{$F_1$ scores of the different algorithms with the different sensors.}
\label{tbl:comparison}
\begin{tabular}{ccccccccc}
\hline
                    & \acrshort{imu}   & \acrshort{gyr}   & \acrshort{gsr}   & \acrshort{hrv}   & \acrshort{imu}+\acrshort{gyr} & \acrshort{gsr}+\acrshort{hrv} & ALL  \\
                    \hline
XGBoost             & \textbf{0.779} & \textbf{0.751} & \textbf{0.723} & \textbf{0.741} & \textbf{0.812}   & \textbf{0.748}  & \textbf{0.820} \\\hline
LGBM                & 0.773 & 0.716 & 0.707 & 0.585 & 0.808   & 0.718  & 0.811 \\\hline
Decision Tree       & 0.707 & 0.726 & 0.698 & 0.686 & 0.741   & 0.703  & 0.771 \\\hline
Logistic Regression & 0.693 & 0.704 & 0.708 & 0.578 & 0.723   & 0.717  & 0.753 \\\hline
\end{tabular}
\end{table}

Note that the \acrshort{gsr} sensors were placed in three positions: left shoulder, right-hand fingertips, and left axilla. As shown, the \acrshort{gsr} feature from the left axilla location is the most indicative. Additionally, the \acrshort{imu} and \acrshort{gyr} sensors were placed on the wrist and elbow, and the data were enriched with a first derivative (``jerk''). Table \ref{tbl:Feature_importance} suggests that the ``jerk'' feature plays a significant role as an input for almost all algorithms. For the \acrshort{hrv} class, the high frequency (HF) feature is the most influential in most algorithms.

Table \ref{tbl:comparison} concludes our model development results, presenting a comparison of the accuracy achieved by the four algorithms. The columns represent different sensor subsets that were used as input for the classification algorithm.
% \begin{table}[ht]
% \begin{tabular}{||c|c|c|c|c|c|c|c||}
% \hline
%                     & IMU   & GYR   & GSR   & HRV   & IMU+GYR & GSR+HRV & ALL  \\
%                     \hline\hline
% XGBoost             & 0.779 & 0.751 & 0.723 & 0.741 & 0.812   &  0.728  & 0.82 \\\hline
% LGMB                & 0.773 & 0.716 & 0.707 & 0.585 & 0.808   & 0.728   & 0.77 \\\hline
% Decision Tree       & 0.707 & 0.726 & 0.698 & 0.686 & 0.711   &    & 0.73 \\\hline
% Logistic Regression & 0.693 & 0.704 & 0.738 & 0.578 & 0.703   &    & 0.72 \\\hline
% \end{tabular}
%     \caption{$F_1$ scores of the different algorithms with the different sensors}
%     \label{tbl:comparison}
% \end{table}
 % It is apparent in the result of our sensitivity analysis that the motion component (i.e., \acrshort{imu}+\acrshort{gyr}) is a prominent input, to the extent that motion alone yields a fairly good model.
As observed, the combination of the \acrshort{imu} sensors and \acrshort{gyr} sensors yields an average approximation of $97\%$ for the optimal $F_1$ score (with all features), while \acrshort{gsr} and \acrshort{hrv} achieve an average of $91\%$ of the total $F_1$ score. Furthermore, the feature extraction for \acrshort{gsr} is based on an auto-encoder algorithm, which results in a relatively high time complexity. This suggests that we %may decrease the number of calculated features (or sensors) without a major compromise in the accuracy, while significantly improving the time complexity. 
can potentially reduce the number of calculated features or sensors without significantly compromising accuracy, while significantly improving the time complexity.

%According to the Flow theory, the fairly good model achieved by the motion components (i.e., \acrshort{imu}+\acrshort{gyr}) may be attributed to the uniqueness of motor activities. Given the inclusion of highly trained musicians, \acrshort{eng} is attained strictly when the complexity of the task is high, which is likely to be manifested by high motion volatility.
According to the Flow theory, the relatively good model achieved by the motion components (i.e., \acrshort{imu} + \acrshort{gyr}) can be attributed to the distinctive nature of motor activities. Considering the inclusion of highly trained musicians, \acrshort{eng} is experienced only when the task complexity is high, which is likely to be reflected in high motion volatility.

For replicability of our results, a lightweight version of the trained XGBoost model ($F_1=0.801$) is available at \url{github.com/AnonymousAuthors}
% \url{http://github.com/IBM/SAX/CONBOTS/ICMI22}
We have also included an XGBoost wrapper that allows users to experiment with model training using their own datasets.

\section{Conclusion and Future work}\label{sec:conclusion}

Our work presents a model for the continuous classification of \acrshort{eng}. We employed an experimental environment with skilled participants who played the Violin while sensor data were collected. To employ supervised learning, we annotated the data by developing a behavioral model. Subsequently, we increased the data density by utilizing interpolation. The labels served as an input for training a machine learning model that predicts the musician's \acrshort{eng} in near real-time.
We discussed in detail the data preparation methodology for the machine learning model, analyzed the results, and performed a sensitivity analysis with respect to the different sensor types.
% The best algorithm achieves an average $F_1$ score of $0.82$ and has ongoing responsiveness of $15$ sec.

We acknowledge that our model may still be susceptible to further testing of its performance in other realistic settings. As a first step in this direction, we applied our model to the recordings in our experiment to produce a prototypical engagement footprint for each excerpt. Our aim is to employ a panel of experienced musicians to validate its faithfulness to the actual musical temperament of each excerpt.

We also note that our setup excluded a leave-1-user-out configuration since we assumed that, in learning contexts such as in our case (i.e., violin practice sessions), it is plausible to capture a baseline recording from each user. However, in the future, we may relax this assumption and attempt a more generalizable configuration.

In the future, we intend to deploy the developed model into a broader teacher-learner interaction system. This system will enable the ongoing tracking of \acrshort{eng} as part of the information about the learner's state. It will be particularly useful as learners gradually improve and progress along the motor learning activity. The measurement of \acrshort{eng} will serve as a key feedback mechanism for the instructor, who has control over the learning pace and rewards allotted to the student's performance.

Furthermore, our future plans include employing the instrumentation for other motor learning activities, such as handwriting and drum playing in children. This will provide an opportunity to test the applicability of the developed instrumentation with different types of motor learning activities, ensuring its robustness.

\bibliographystyle{ACM-Reference-Format}
\bibliography{biblo}

\appendix

\section{Appendix}
\begin{figure}
    \centering
    \includegraphics[width=0.95\linewidth]{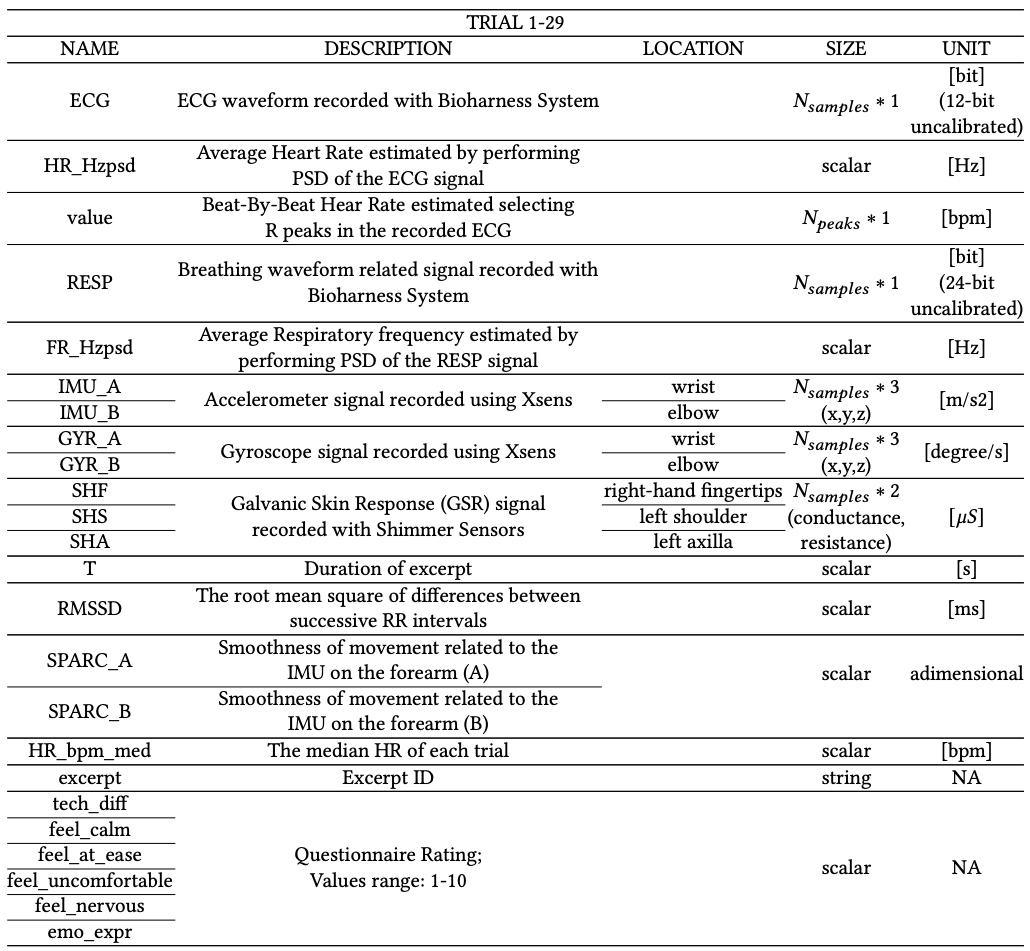}
    \caption{Experimental measurements}
    \label{fig:my_label}
\end{figure}

\end{document}